\documentclass[10pt,letterpaper]{article}
\usepackage[top=0.85in,left=1in,footskip=0.75in]{geometry}
\usepackage[utf8]{inputenc}
\usepackage{cite}
\usepackage{nameref,hyperref}
\usepackage[left]{lineno}
\usepackage{titlesec}
\usepackage{lineno}
\usepackage{upgreek}
\usepackage{ragged2e}
\usepackage{setspace} 
\usepackage{changepage}
\usepackage[aboveskip=1pt,labelfont=bf,labelsep=period,singlelinecheck=off]{caption}
\usepackage{lastpage,fancyhdr,graphicx}
\usepackage{epstopdf}
\usepackage{upgreek}
\usepackage{lipsum}
\usepackage{color}
\usepackage{graphicx}

\justifying
\begin{document}
\vspace*{0.35in}

\begin{center}
\Huge\textbf{Relayed-loop optical scan amplification}
\newline
\\
\large
Harishankar Jayakumar\textsuperscript{1}\dag,
Christopher Warkentin\textsuperscript{1}\dag,
Deano Farinella\textsuperscript{1},
Samuel Stanek\textsuperscript{2},
Zachary L Newman\textsuperscript{1},
Runze Liu\textsuperscript{2},
Aaron Kerlin\textsuperscript{1,*}
\\
\bigskip
{1} Department of Neuroscience, Center for Magnetic Resonance Research, University of Minnesota, 2021 6th Street SE, Minneapolis, MN 55455, USA
\\
{2} Department of Electrical and Computer Engineering, University of Minnesota, 200 Union Street SE, Minneapolis, MN 55455, USA
\\
\dag These two authors contributed equally to this work.
\\
\bigskip
* akerlin@umn.edu

\end{center}

\section*{Abstract}
Many modern sensing, processing, and fabrication technologies depend upon the sequential scanning of laser light. Due to inertial, thermal, and electrical limitations, the speed of a typical laser deflector is inversely related to its optical invariant and therefore the number of resolvable spots it can address. Passive optical systems have been developed that can effectively double the optical invariant and maximum throughput of a single deflector. We present a method for increasing the effective optical invariant beyond 2$\times$ by repeated relaying of the deflected beam onto the deflector in an optical loop created within a cavity. We use this Relayed-Loop Optically Amplified Deflection (ReLOAD) approach to accomplish 8$\times$ amplification of electro-optical (EO) deflection, a deflection technology with unparalleled speed, but limited optical invariant. Using ReLOAD, we demonstrate 10 kHz frame rate imaging, 1 MHz line scan rate, and $\mu$s step times across an addressable space well beyond the capabilities of typical electro-optical deflectors.  

\noindent\rule{\textwidth}{0.5pt}
\newpage
High-speed laser deflectors find a wide range of applications, including imaging, mapping, manufacturing, information storage, and computation \cite{marshall_handbook_2012}.  The utility of a deflector across these domains is strongly influenced by both its operating speed and the number of resolvable spots it can address, as these parameters set limits on the field-of-view (FOV) at a given optical resolution. In certain use cases, both high speed and high number of resolvable spots are critical in order to record \cite{demas_high-speed_2021} or control \cite{schindler_quantum_2013} fast processes occurring across many locations. However, inertial, thermal, and electrical constraints typically impose a trade-off between the address rate and the number of resolvable spots. This trade-off is apparent for mechanical deflectors, such as galvanometers and resonant mirrors, as well as non-mechanical deflectors, such as acousto-optical deflectors (AODs) and electro-optical deflectors (EODs). Recently, some optical systems have been developed that circumvent this trade-off, achieving a 2$\times$ amplification of the effective throughput (\textit{i.e.,} resolvable spots per second) of reflective deflectors \cite{xiao_scan_2021,hebert_improving_2024,zolfaghari_cascaded_2024, nomura_scanner_2004}.
\begin{figure}[h!]
    \centering
    \includegraphics[width=0.75\columnwidth]{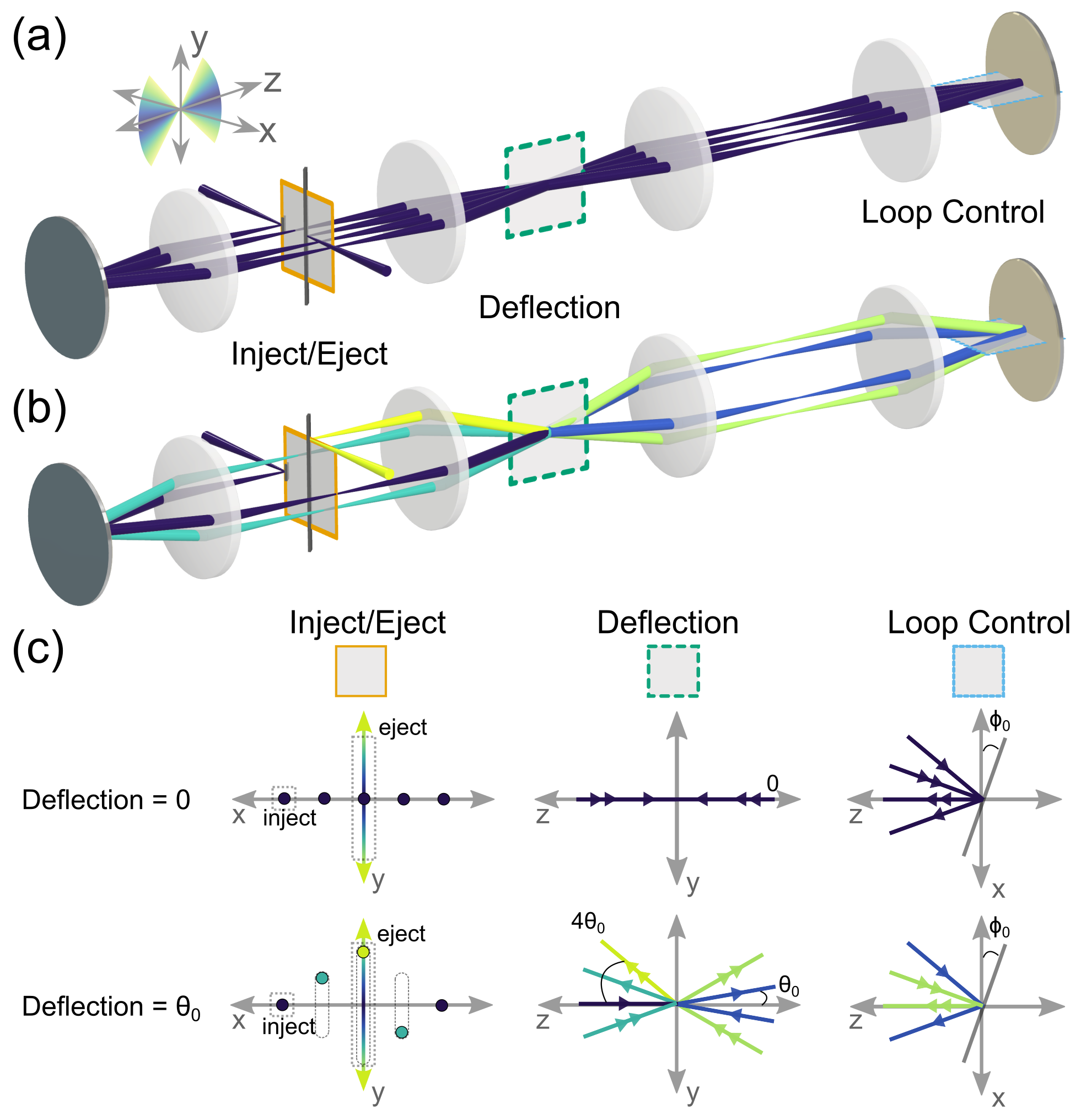}
    \caption{Illustrations of transmissive ReLOAD configured for 4$\times$ amplification. (a) Optical path of a laser beam with no deflection applied at the transmissive deflector. (b) Identical to (a), but with deflection of $\theta_0$ applied at the deflector. (c) \textit{xz} cross-section at the inject/eject mirror (orange solid), \textit{yz} projection at the deflector (teal dashed) and \textit{xz} projection at the tilted end mirror (light blue dotted) shows the beam paths traversed by an undeflected (top panels) and deflected (bottom panels) beam. Black dashed lines indicate locations of the inject and eject mirrors and the dotted lines indicate the extent of the bidirectional deflection. The number of arrow heads indicate the loop number. }
    \label{fig:reload_concept}
\end{figure}

To achieve scan amplifications greater than 2$\times$, we have designed a method that repeatedly relays the deflected beam back to the deflector in an optical loop within a cavity \cite{kerlin_relayed_2024} (Fig. \ref{fig:reload_concept}), which we call Relayed-Loop Optically Amplified Deflection (ReLOAD). Here, we focus on an implementation of ReLOAD for amplifying a transmissive non-mechanical deflector. However, ReLOAD is a versatile \textit{N}-fold deflection amplification scheme that is also compatible with reflective mechanical deflectors (Fig. \ref{fig:mech_reload}, see \hyperref[sec:Demonstration Video]{Visualization 1}).  Conceptually, ReLOAD requires a cavity with a loop dimension (\textit{xz} Fig. \ref{fig:reload_concept}(a)) that sets the number of interactions of light with the active deflector before ejection from the cavity. This loop dimension is orthogonal to the active deflection dimension (\textit{yz}), allowing the amplification factor to be set independently. In our physical implementations of ReLOAD, an optical cavity is realized with afocal relays and a deflector placed at the afocal plane. For transmissive deflectors, the deflector is placed at the common afocal plane of two sequential afocal relays (Fig.\ref{fig:reload_concept}). Light enters the cavity \textit{via} a small inject mirror placed at the focal plane of the cavity. A loop control mirror on one end of the cavity is tilted at an angle $\phi_0$ (Fig. \ref{fig:reload_concept}(c)). The applied angle (2$\phi_0$) causes the beam to follow a converging figure-eight trajectory, with each subsequent loop spanning a smaller angle in the loop dimension. In this way, the beam interacts with the deflector multiple times -- accumulating additional angle ($\theta_0$) along the deflection axis with each pass (Fig. \ref{fig:reload_concept}(b)), so that the deflection angle after \textit{N} passes is \textit{N$\theta_0$}. The looping beam exits the cavity by reflecting off an eject mirror placed at the focal plane (Fig. \ref{fig:reload_concept}(c)). The number of loops is determined by the tilt angle ($\phi_0$) of the loop control mirror, with the maximum number of loops limited by the width of the eject mirror, as well as other practical constraints (\textit{e.g.,} acceptance angle of the system, cumulative aberrations, and cumulative power loss).    

\begin{figure}[h!]
    \centering
    \includegraphics[width=0.75\linewidth]{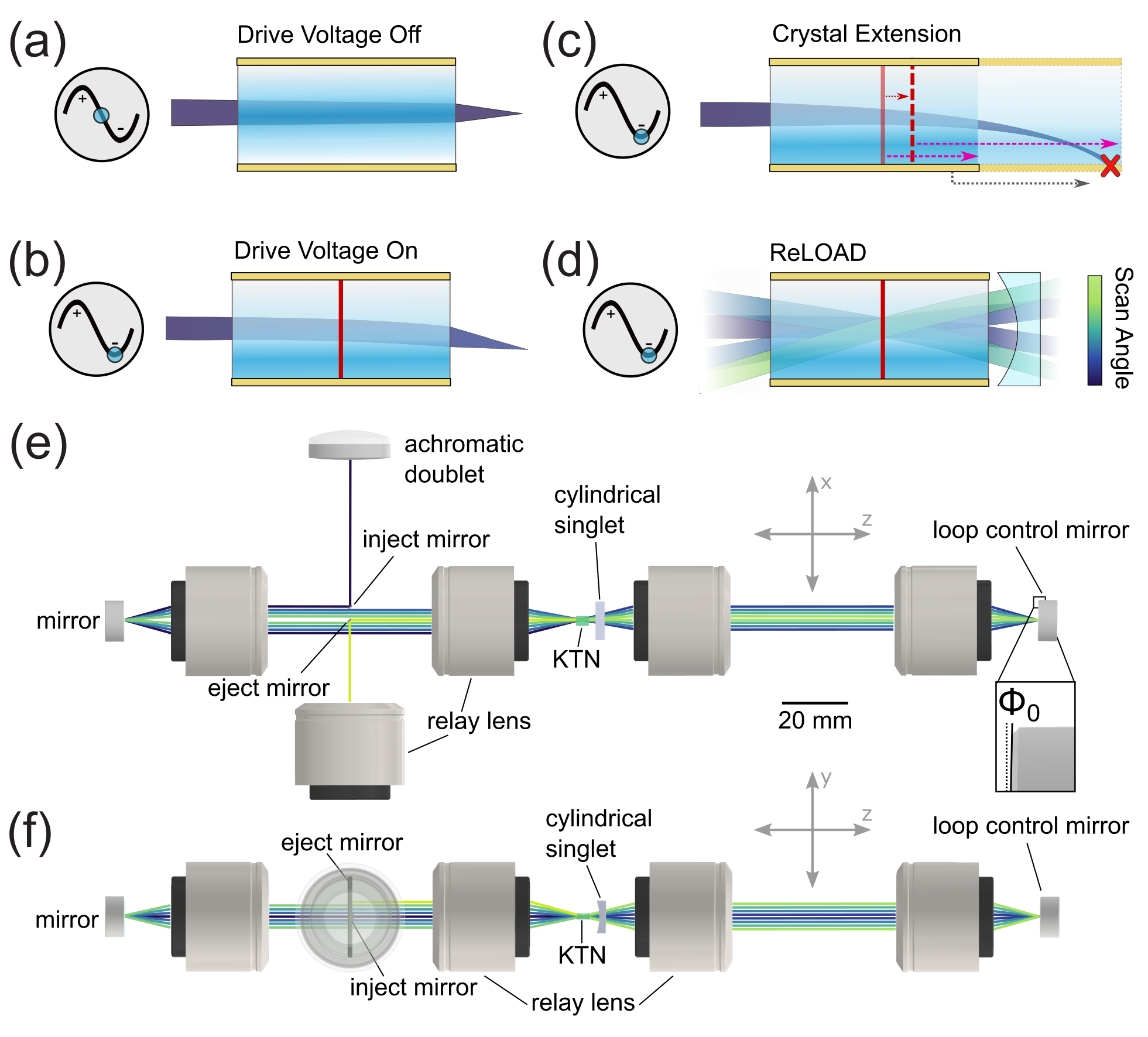}
    \caption{Illustrations and schematics of EO-ReLOAD with a KTN EOD deflector. (a) A cylindrical convex gradient index (GRIN) formed within the charged KTN (b) Drive voltage shifts the optical axis of the GRIN, deflecting the beam. Red line: pivot plane of the deflected beam (c) Increasing deflection angle by extending interaction length is limited by vignetting at the sidewalls and increasing capacitance.  Dashed red line: new pivot plane. (d) Beam paths with 8$\times$ ReLOAD amplification. (e) Scale view of the loop dimension (\textit{xz}) of the optical cavity with the KTN-cylindrical lens pair at the common afocal plane of two relays. Inset: tilt angle of the loop control end mirror. (f) Deflection dimension (\textit{yz}) of the cavity. 
    }
    \label{fig:reload_KTN}
\end{figure}

To demonstrate the utility of ReLOAD, we applied it to electro-optical deflection (Fig. ~\ref{fig:reload_KTN}), a deflection technology with unparalleled speed, but low optical invariant and number of resolvable spots. We constructed an EOD based on the perovskite crystal, potassium tantalate niobate (KTN, KTa$_{1-x}$Nb$_x$O$_3$). In its paralectric phase, the large quadratic electro-optic coefficient and a high dielectric constant of KTN \cite{chen_light_1966} yields an unusually large deflection angle per unit of electric field compared to other EOD technologies \cite{zhu_advanced_2018,romer_electro-optic_2014}. KTN-based EODs can operate at up to MHz rates \cite{shang_higher_2023, zhu_three_2016}. However, EODs designed for high-speed ($>$50 kHz line rate) operation can address a very limited number of resolvable spots due to a combination of issues, including clear aperture of the crystal, EOD capacitance, and electroelastic damage \cite{zhu_photon_2018, zhu_three_2016}. Commercial KTN-based deflectors can address a maximum of $\approx$32 resolvable spots at a 200 kHz line rate and only $\approx$4 resolvable spots at 1 MHz line rate \cite{farinella_two-dimensional_2024}.  

Our custom-built KTN-based EOD (Fig. \ref{fig:holder}) contains a 4 mm $\times$ 3.15 mm $\times$ 1.2 mm KTN crystal plated with layered titanium/platinum/gold electrodes (NTT Advanced Technology Corp.). A broadband magnesium fluoride anti-reflective coating was applied to transmissive faces \textit{via} electron beam evaporation. Copper interfaces provided electrical contact with the crystal and thermal contact with thermoelectric modules that maintained the KTN at 5 °C above its Curie temperature (T\textsubscript{c} = 23 °C), ensuring operation in the paraelectric phase (see \hyperref[sec:Supporting Information]{Supporting Information}). The deflector was operated in a space-charge mode \cite{miyazu_new_2011} by applying a temporary 400 VDC bias to fill electron traps in the crystal. According to Gauss's law, this charge results in a 1-dimensional parabolic gradient refractive index profile (cylindrical convex GRIN lens) orthogonal to the electrodes \cite{imai_analyses_2017}, with a back focal length of $\approx$ 32-40 mm (Fig. ~\ref{fig:reload_KTN}(a)). Under our operating conditions, charges remained trapped in the crystal for hours \cite{stanek_optical_2024}.

Variable voltage can be applied across the KTN to shift the GRIN lens (Fig. \ref{fig:reload_KTN}(b)), causing the light beam passing through the crystal to deflect. The deflection angle of the beam for a single-pass through the KTN is given by 

\begin{equation}
    \theta_{KTN}=\frac{n^3Vg_{11}\varepsilon\rho L}{D}
    \label{eq:KTN_angle}
\end{equation} 

\noindent where, $n$ is the refractive index, $V$ is the voltage applied across the crystal, $g_{11}$ is the quadratic electro-optic coefficient, $\varepsilon$ is the permittivity, $\rho$ is the charge density, and $L$ and $D$ are the length and electrode separation distance of the KTN, respectively \cite{imai_analyses_2017}. For a single-pass through the EOD, the acceptance angle and maximum number of resolvable spots is limited by the aperture (\textit{i.e.,} electrode gap), length of the crystal along the optical axis, and the electric-field-induced damage threshold of the crystal. Increasing the aperture size makes crystal charging and driving more difficult \cite{zhu_photon_2018}. Extending crystal length will increase the deflection angle but decrease the acceptance angle due to vignetting at the crystal walls (Fig. \ref{fig:reload_KTN}(c)). Using ReLOAD, the large acceptance angle of short EODs can be fully exploited (Fig.\ref{fig:reload_KTN}(d)).

Our implementation of ReLOAD for the KTN-based EOD (EO-ReLOAD) is shown in Fig. \ref{fig:reload_KTN}(e, f). The 8-$f$ EO-ReLOAD cavity consisted of two 4-$f$ relays, formed by four telecentric scan lenses (LSM03-BB, Thorlabs), and two dielectic end mirrors. The EOD was placed at the common afocal plane of the two relays, conjugating its pivot plane to the end mirrors. The GRIN lens formed inside the KTN crystal also focused the beam along the deflection axis. To compensate for the focus of the beam by this GRIN lens, a cylindrical concave lens ($f$ = -30 mm) was placed on one side of the deflector to maintain a collimated beam \cite{imai_analyses_2017, farinella_two-dimensional_2024} (Fig. \ref{fig:reload_KTN}(d, e, \textit{f})). At the focal plane of one relay, the light beam entered the cavity as a converging beam with its waist centered on the off-axis inject mirror.  Light was ejected from the cavity by a high-aspect-ratio rectangular pick-off mirror placed on-axis within the same focal plane. Deflection angle accumulated over $N$ repeated passes through the EOD.

In the case of an ideal ReLOAD system with folded 8-$f$ relays (Fig. (\ref{fig:reload_concept}c)), the maximum amplification ($N_{max}$) is given by 

\begin{equation}
    N_{\mathrm{max}} =
    \left\lfloor
    \min\left(
        \frac{\Theta(d)}{\theta_{0,\mathrm{max}}},
        \frac{\Phi(d)}{\phi_{0,\mathrm{min}}}
    \right)
    \right\rfloor
\end{equation}

\noindent where $\Theta$ and $\Phi$ are the system acceptance angles along the deflection dimension and loop dimension, respectively, given $1/e^2$ input beam diameter $d$. In our case $\theta_{0,max}$ is the maximum deflection angle of the deflector and cylindrical lens combination independent of ReLOAD and $\phi_{0,min}$ is the minimum angle of the loop control mirror. Furthermore, if the inject mirror and the eject mirror are located in the same plane, $N_{max}$ must be even. To eject 99\% of a Gaussian beam at the focal plane using a mirror of ideal width (see \hyperref[sec:Supporting Information]{Supporting Information}), $ \phi_{0,min}$ is given as

\begin{equation}
    \phi_{0,\min} = 0.5 \arctan \left( \frac{6.4 \lambda}{\pi d}\right)
    \label{eq:min_angle}
\end{equation}

\noindent where $\lambda$ is the wavelength of the light. For our EO-ReLOAD module, where $\Theta\left(d\right)/\theta_{0,max}<<\Phi\left(d\right)/\phi_{0,min}$, the acceptance angle of the relay lenses determined $\Theta(d)$, and $N_{max}=14$ (see \hyperref[sec:Supporting Information]{Supporting Information}). To maintain reasonable power throughput, we configured the module for an amplification of $N = 8$ rather than $N_{max}$. The maximum number of resolvable spots addressable by our deflector in ReLOAD is given by

\begin{equation}
    RS_{FWHM}=\frac{2Nd\tan{\theta_{def,max}}}{0.75 \lambda} +1
    \label{eq:res_spots}
\end{equation}

\noindent where $\theta_{def,max}$ is the maximum deflection angle of the deflector (\textit{i.e.} $\theta_{KTN,max}$ in this implementation) independent of ReLOAD. Based on the above equations, ideal diffraction-limited performance of our 8$\times$ EO-ReLOAD module should result in $RS_{FWHM}=177$ resolvable spots (see \hyperref[sec:Supporting Information]{Supporting Information}). 

\begin{figure}[h!]
    \centering
    \includegraphics[width=0.75\linewidth]{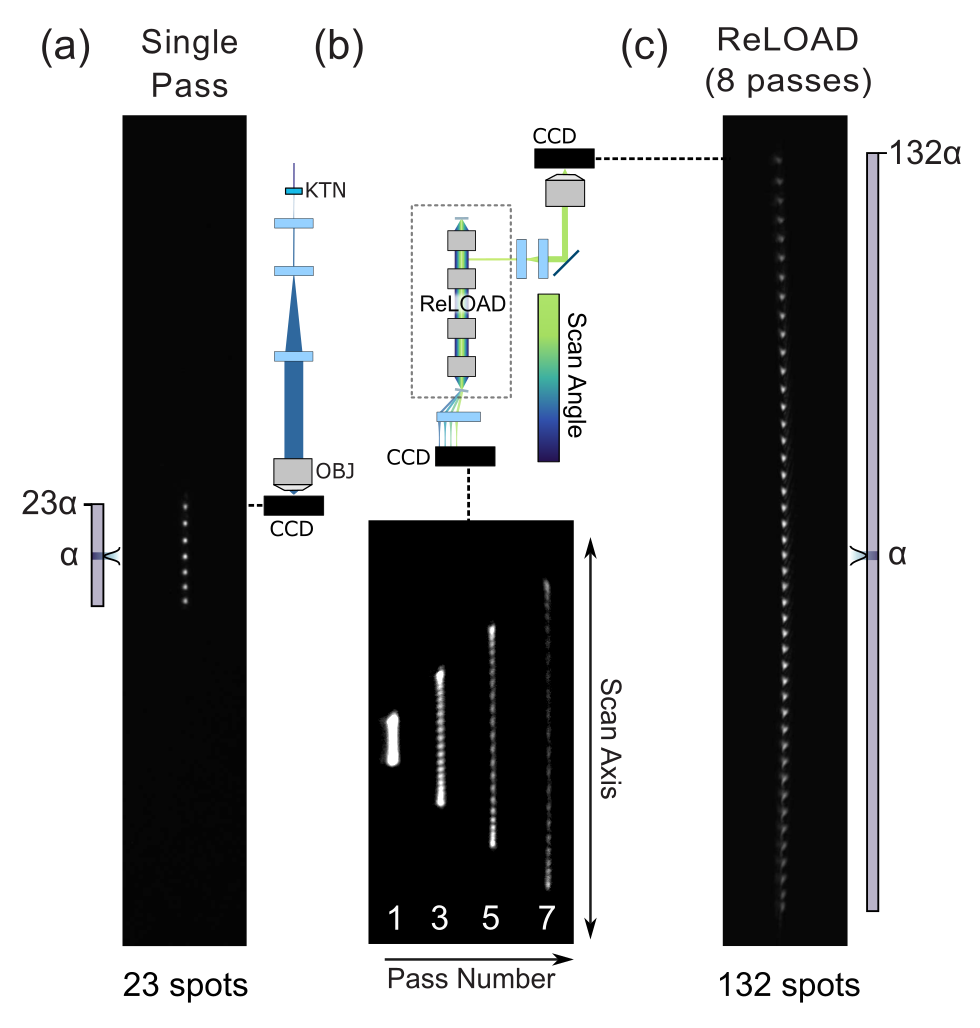}
    \caption{Stroboscopic imaging of deflector performance. (a) Combined image of stroboscopic images of spots spanning the range of the KTN deflector outside ReLOAD. Adjacent illustration: imaging configuration. (b) Image of scan range linearly increasing with pass number collected from behind an end mirror of the EO-ReLOAD module. (c) Same as (a), but for the output of the EO-ReLOAD module. Gray bars: range of deflection. Blue shaded region within the bar ($\alpha$): average spot FWHM.}
    \label{fig:reload_KTN_results}
\end{figure}

Stroboscopic imaging was used to evaluate the output of the EOD operated independently of ReLOAD and the output of the complete EO-ReLOAD apparatus (Fig. \ref{fig:reload_KTN_results}(a, c)). The EOD was driven by a 100 kHz (200 kHz line rate), 592 $\mathrm{V_{pp}}$ sinusoidal signal generated by a signal generator that was amplified 100$\times$ by two bridged amplifiers (Advanced Energy, Trek 2100HF). The deflection system output was focused onto a CMOS camera. Synchronization pulses from the signal generator triggered laser light emission (Edinburgh Instruments, EPL-980; 970 nm, 60 ps) at a set phase with respect to the EOD drive, enabling sequential collection of spot images across the scan range (see \hyperref[sec:Supporting Information]{Supporting Information}). Independent of ReLOAD (\textit{i.e.}, single-pass), the EOD scanned a maximum of 23 resolvable spots (FWHM) (Fig. \ref{fig:reload_KTN_results}(a)). In the EO-ReLOAD module, configured for 8$\times$ amplification, the EOD scanned 132 resolvable spots (Fig. \ref{fig:reload_KTN_results}(c); Fig. \ref{fig:all_psf_data}), equivalent to a 6$\times$ amplification. To assess the sublinear scaling of resolvable spots in this implementation of ReLOAD, light leaking through one dielectric end mirror was focused onto a CMOS camera to simultaneously image the scan range of sequential loops. The deflection range was observed to increase linearly with the number of loops through EO-ReLOAD (Fig. \ref{fig:reload_KTN_results}(b)), demonstrating that angle amplification occurs as expected. Thus, the sublinear amplification of resolvable spots likely reflects an increase in average spot size ($\approx$ 25\%) caused by cumulative aberrations in the system. Spot intensity also declined by as much as 76\% from the center to the edge of the defection range (Fig. \ref{fig:reload_KTN_results}(c); Fig. \ref{fig:all_psf_data}), due to unanticipated vignetting at the edges of the crystal (data not shown). Optical simulation indicated that this suboptimal performance may reflect tight tolerances on the telecentricity of the scan lenses in ReLOAD, which are tighter than the manufacturing tolerances of the stock lenses used in this implementation. Deviations from telecentricity will shift the beam along the deflection axis at the afocal conjugate planes, resulting in vignetting at the crystal edges. The power throughput of the 8$\times$ EO-ReLOAD module was 11\%. Future implementations of ReLOAD with custom-manufactured scan lenses and optimized anti-reflection coatings should enable angle amplification with significantly higher performance.

\begin{figure}[h!]
    \centering
    \includegraphics[width=0.75\linewidth]{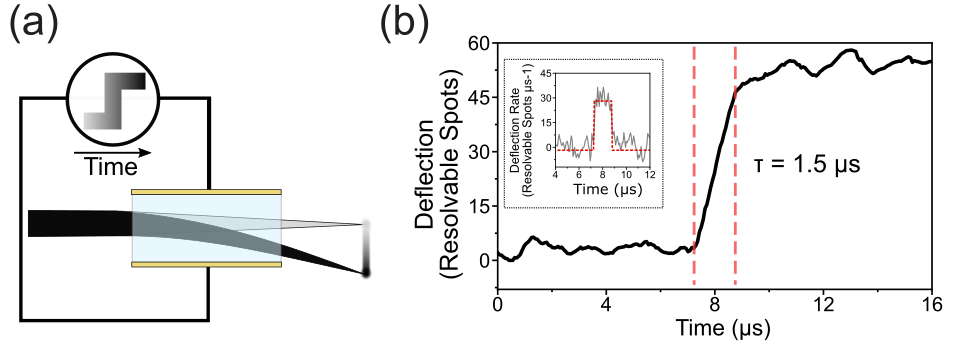}
    \caption{Step response EO-ReLOAD. (a) Illustration of KTN deflector driven with a rectangular wave. (b) Timecourse of deflection (black line) in response to a 160 V step. Dashed red lines: extent of the step period ($\tau$). Inset: Deflection rate (black line) and fit (dashed red line) to estimate the step period.}
    \label{fig:reload_KTN_responsetime}
\end{figure}

In addition to being capable of rapid sinusoidal scanning, EODs are among the fastest deflectors capable of random-access targeting of light \cite{romer_electro-optic_2014}. KTN-based deflection has been reported with nanosecond step times \cite{chao_high_2017}. However, the limited gain-bandwidth product of high-voltage drive electronics typically restricts KTN-based EODs to longer step times. Stroboscopic imaging of the response of our EOD to a 160 V rectangular signal indicated a step time of 1.5 $\mu$s across a distance equivalent to 50 resolvable spots within 8$\times$ EO-ReLOAD (Fig. \ref{fig:reload_KTN_responsetime}). A similar step time was observed in the voltage output of the drive electronics when driving a 2 nF capacitor in place of the EOD ($C_{EOD}$ = 1.9 nF; charged, 28 °C), confirming that the step time reflects the limitations of the commercial drive electronics. By amplifying EOD deflection, ReLOAD could reduce the crystal dimensions required to access a fixed number of resolvable spots, effectively decreasing EOD capacitance and enabling faster operation. 

\begin{figure}[h!]
    \centering
    \includegraphics[width=0.75\linewidth]{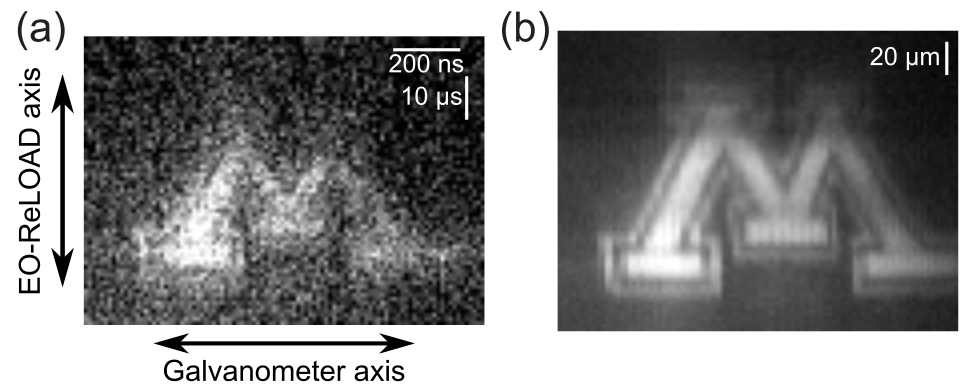}
    \caption{Kilohertz reflectance microscopy with EO-ReLOAD. (a) Single frame in linear time acquired at 1 MHz line rate and 10 kHz frame rate. Pixel dwell time: 12.5 ns.  (b) Average of 100 frames transformed to linear space.}
    \label{fig:reload_microscopy}
\end{figure}

The speed of laser scanning microscopy has become increasingly important in the biological sciences \cite{wu_speed_2021}. To demonstrate how ReLOAD can enable faster imaging, we integrated the EO-ReLOAD module into a laser scanning reflectance microscope (see \hyperref[sec:Supporting Information]{Supporting Information}, Fig. \ref{fig:mult_ref_scope}(b)). Output of the EO-ReLOAD module was relayed to a mirror galvanometer that scanned the orthogonal lateral dimension. The EOD and galvanometer were driven with phase-locked sinusoidal signals, with the EOD at a 100$\times$ harmonic of the galvanometer frequency (500 kHz and 5 kHz, respectively). At 500 kHz, the drive electronics generated a smaller Vpp, limiting EO-ReLOAD output to 50 resolvable spots (\hyperref[sec:Supporting Information]{Supporting Information}, Fig. \ref{fig:all_psf_data}(f)). Light (960 nm) was provided by an 80 MHz infrared laser commonly used for nonlinear microscopy. Detector signals were reconstructed and displayed in real-time using commercial microscope control systems (MBF Biosciences, ScanImage). The imaging system was capable of imaging a microscopic metal mask in the shape of a university logo at 10 kHz bidirectional frame rate and 1 MHz bidirectional line rate (Fig. \ref{fig:reload_microscopy}). This performance is comparable to the MHz line rates of passive scanners recently developed for two-photon (2P) laser scanning microscopy \cite{lai_high-speed_2021}, but EO-ReLOAD offers the advantage of not necessitating the use of $\mu$J-class femtosecond lasers or custom acquisition systems that are required for some pulse-splitting schemes. We note that, due to the excitation confinement inherent to two-photon absorption, this implementation of EO-ReLOAD would be expected to scan 185 2P-resolvable spots at 200 kHz line rate and 70 2P-resolvable spots at 1 MHz line rate.

The development of systems that double scan angles \cite{xiao_scan_2021,hebert_improving_2024,nomura_scanner_2004} showed that passive optical systems can increase deflector throughput. ReLOAD provides a versatile method for amplifying deflection $N$-fold and should enable improvements to the scan range, throughput, and access times of a wide range of transmissive and reflective deflectors. We note that during preparation of this manuscript, a preprint was posted that presented a similar method to amplify the scan angle of reflective deflectors \cite{yu_non-inertial_2025}. Among deflection technologies, we expect ReLOAD to have the most transformative effect in the case of EODs. Using a commercial deflector, we previously demonstrated electro-optical 2P imaging of neurons \textit{in vivo} \cite{farinella_two-dimensional_2024}. However, the FOV was too small to be of practical use.  EO-ReLOAD could enable future electro-optical 2P microscopes to access hundreds of resolvable spots at exceptional speeds. The speed of EO-ReLOAD may also prove beneficial in applications such as improving throughput in 3D nanomanufacturing \cite{saunders_ultrafast_2023} or reducing gate latency in trapped-ion quantum computing \cite{schwerdt_scalable_2024}. To achieve these impacts, further work will be required to develop higher throughput and smaller footprint implementations of EO-ReLOAD.

\newpage
\section*{Supporting Information}
\phantomsection
\label{sec:Supporting Information}
\setcounter{figure}{0}
\renewcommand{\thefigure}{S\arabic{figure}}

\section*{Reflective mechanical ReLOAD}
\addcontentsline{toc}{section}{Mechanical ReLOAD}

\subsection*{Concept}
\phantomsection
\addcontentsline{toc}{subsection}{Concept}
A CAD schematic of the reflective mechanical ReLOAD design is shown in Figure~\ref{fig:mech_reload}(a,b). As described in the main text, a fully mechanical version of ReLOAD is an implementation in which the deflector is any kind of reflective inertial scanner (\textit{e.g.,} mirror galvanometer, resonant mirror, polygon mirror, MEMS mirror). Figure~\ref{fig:mech_reload}(c) shows the implementation of this mechanical ReLOAD design using a resonant mirror deflector (EOPC, SC-30-16K). The upper-right inset of Figure~\ref{fig:mech_reload}(c) shows a composite image of the amplified scan range of 4 subsequent loops through this system along the loop axis at the \textit{z}-position of the inject lens. The images were extracted from a video that was collected as described in \hyperref[sec:Demonstration Video]{Demonstration Video}.

\subsection*{Optical Path Details}
\begin{figure}[h!]
\centering
\fbox{\includegraphics[width=0.9\linewidth]{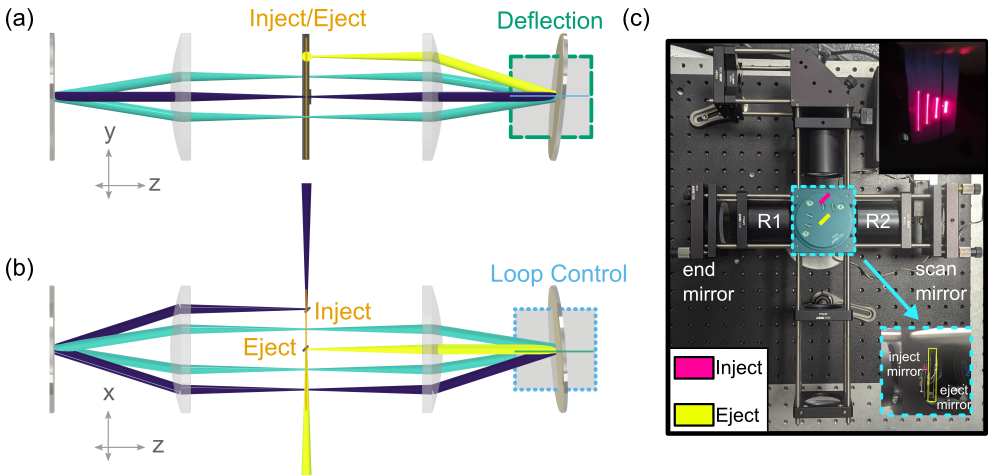}}
\caption{Reflective mechanical ReLOAD. (a) \textit{yz} projection showing dynamic deflection. (b) \textit{xz} projection visualizing static loop control. Deflection is along the \textit{y}-axis, while number of loops is determined by the \textit{x}-axis deflector angle. (c) Photograph of the mechanical ReLOAD equipped with a resonant mirror deflector.Upper-right inset shows a composite image of loops 1-4 in this implementation of mechanical ReLOAD.}
\label{fig:mech_reload}
\end{figure}

\addcontentsline{toc}{subsection}{Optical Path Details}
In reflective mechanical ReLOAD, a laser is focused onto a small inject mirror (Edmund Optics, \#86-621), which redirected the beam towards a scan lens (R1). In this implementation, the compound scan lenses consisted of 3$\times$  achromatic doublets (Thorlabs, AC508-150-B). The inject mirror was located at the outer edge of the field-of-view (FOV) of the scan lens as seen in Figure ~\ref{fig:mech_reload}. The scan lens was placed one focal length away from the inject mirror, and an end-mirror (Thorlabs, PF20-03-M01) was placed one focal length beyond the scan lens. After reflection from the end-mirror, the beam was refocused by the same scan lens, and re-collimated by an identical scan lens (R2), placed 2 focal lengths away from the original. A scan mirror (EOPC, SC-30-16K) was placed at a plane conjugate to the end mirror and located one focal length away from this second scan lens, with the two relay lenses constituting a 4-\textit{f} system. The scanning mirror was placed at an angle to the optical axis of the 4-\textit{f} system along the loop axis (\textit{x}-axis), such that the reflected beam takes a different path through the system closer to the center of the FOV of the scan lens in the inject/eject plane. The angle imparted along the loop axis determines how many reflections are required to passively eject the beam from ReLOAD, which occurs when the beam encounters a rectangular eject mirror located at the center of the scan lens FOV along the loop axis. In the orthogonal scan axis (\textit{y}-axis), each time the beam encounters the scanning mirror, a nearly identical deflection occurs. In this way, the scanning mirror deflection is amplified by the number of loops traversed before reaching the center of the FOV, and can be adjusted by changing the loop axis angle of the scanning mirror.

\subsection*{Demonstration Video}
\phantomsection
\addcontentsline{toc}{subsection}{Demonstration Video}
\label{sec:Demonstration Video}
A video demonstrating deflection amplification in our mechanical ReLOAD system is shown in \href{https://figshare.com/articles/media/Visualization_1/30180574?file=58138006}{Visualization 1}. For demonstration purposes, the eject mirror was removed to more clearly visualize  amplification. Light from a red laser diode was injected into the system and we translated a narrow piece of card stock across the loop axis at the \textit{z} position of the inject mirror. As the card was translated, the scan range for subsequent loops through ReLOAD was visualized. In this demonstration, 5 loops are observed. 

\section*{KTN Deflector Design}
\begin{figure}[h!]
\centering
\fbox{\includegraphics[width=0.9\linewidth]{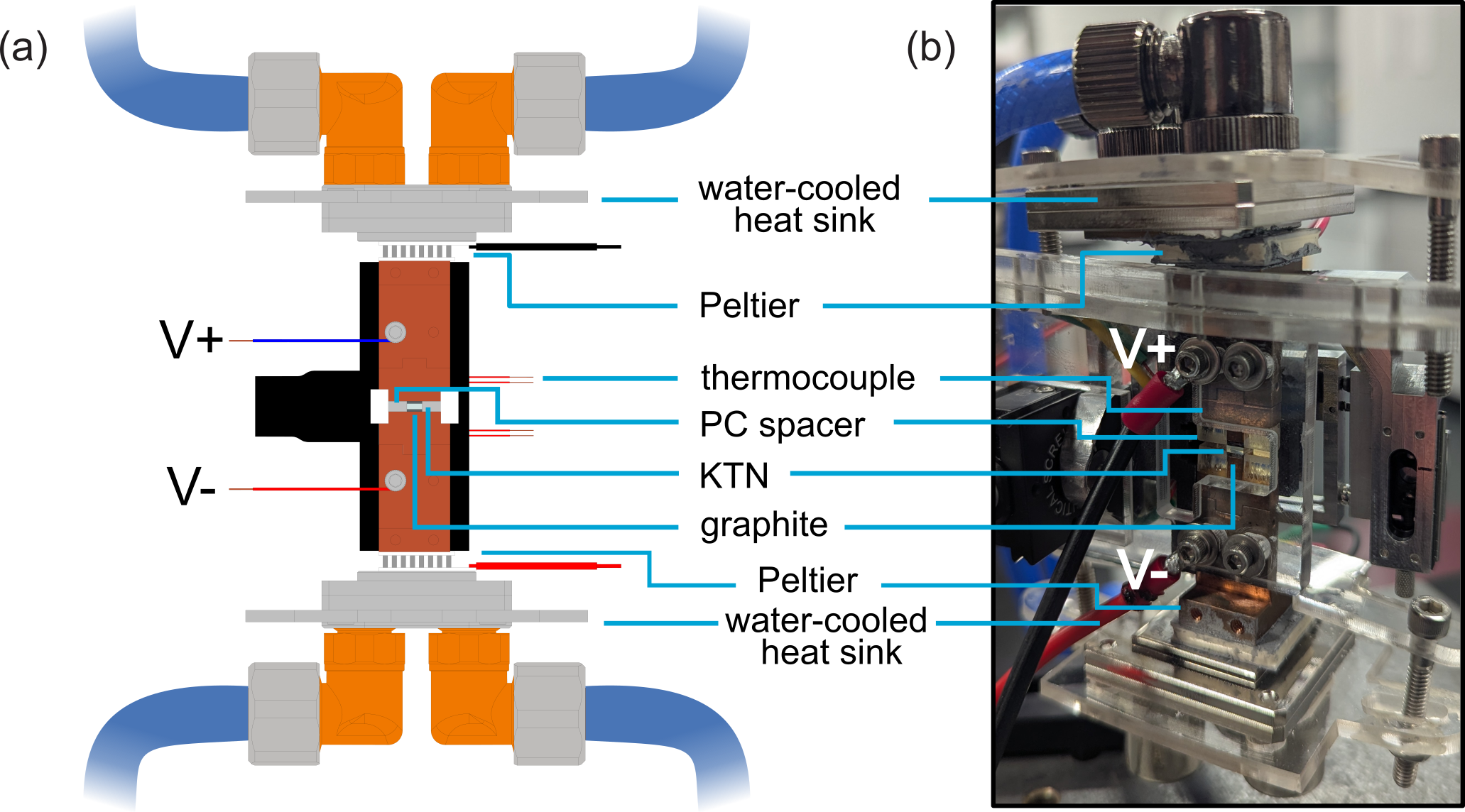}}
\caption{KTN module design. (a) Schematic and (b) photograph of the custom KTN-based deflector module. The module integrated copper electrodes with compressible graphite pads for electrical contact, thermocouples and thermoelectric modules for active temperature regulation, and water-cooled heat sinks for thermal management.}
\label{fig:holder}
\end{figure}

\phantomsection
\addcontentsline{toc}{section}{KTN Deflector Design}
\label{sec:KTN Module Design}
The KTN crystal (1.2 mm $\times$ 3.15 mm $\times$ 4 mm) was purchased from NTT Advanced Technology Corp. and arrived with layered titanium/platinum/gold (Ti/Pt/Au) electrodes plated on both 3.15 mm × 4 mm faces. A broadband magnesium fluoride AR coating ($\approx$177 nm) was also applied to the 1.2 mm $\times$ 3.15 mm faces of the crystal \textit{via} electron-beam evaporation (Varian). The crystal can be charged and driven by applying voltage across these surfaces. The crystal was mounted in a custom module designed for integration with EO-ReLOAD, providing electrical access, mechanical stability, and active temperature control (Fig.~\ref{fig:holder}). Voltage was delivered to the KTN through two custom-milled copper electrodes, each capped with a custom cut compressible graphite pad (McMaster-Carr, 1276N12) to ensure reliable contact with the Ti/Pt/Au-coated surfaces of the crystal. The graphite pads also provided mechanical compliance, accommodating minor crystal flexion that may occur while charging and driving the crystal \cite{imai_anomalous_2015}. Temperature stabilization was achieved through a dual-control scheme. Thermocouples (TC, Thorlabs, THK10) were mounted adjacent to the KTN within each copper electrode, while thermoelectric modules (TEM, McMaster-Carr, 9383N13) were placed at the electrode ends. Each TC/TEM pair was connected to a separate temperature controller (Cell, TDC-1010A), which adjusted the TEM output based on TC feedback to maintain a temperature of $T_C$ + 5 °C. The module was coupled to a liquid-cooling system (Koolance, EX2-1050BK). Heat was dissipated from the backside of the TEMs through cold plates (Koolance, PLT-UN25F). This combined TEM and liquid-cooling approach prevented the development of thermal gradients across the crystal, ensuring stable operation and preventing performance degradation during high-frequency beam deflection \cite{farinella_two-dimensional_2024}.

\section*{Equation Details}
\addcontentsline{toc}{section}{Equation Details}

\subsection*{Equation 2}
\phantomsection
\addcontentsline{toc}{subsection}{Equation 2}
\noindent In Eq.2, we described the the maximum number of loops ($N_{max}$) in the case of an ideal ReLOAD system with folded 8-$f$ relays as

\begin{equation}
    N_{\mathrm{max}} =
    \left\lfloor
    \min\left(
        \frac{\Theta(d)}{\theta_{0,\mathrm{max}}},
        \frac{\Phi(d)}{\phi_{0,\mathrm{min}}}
    \right)
    \right\rfloor
\end{equation}

\noindent For the demonstration of EO-ReLOAD in this experiment, both $\Theta(d)$ and $\Phi(d)$ were determined by the acceptance angle of the relay lenses ($\approx 185$ mrad), $\theta_{0,max}=12$ mrad, and $\phi_{0,min}=1.2$ mrad (see below). Given that the inject and eject mirrors were configured to lie in the same plane and thus $N$ must be even, this gives $N_{max} = 14$. Note that, in other configurations of EO-ReLOAD, $\Theta(d)$ will be determined by the effective acceptance angle of the electro-optical crystal. For KTN-based EODs, analytical derivation of this acceptance angle would be complex, because the pivot point of deflection shifts slightly each loop. Instead, we estimated the maximum effective acceptance angle of the KTN in simulation (with ideal relay lenses) to be $\approx$240 mrad.

\subsection*{Equation 3}
\addcontentsline{toc}{subsection}{Equation 3}

In Eq.3, we described the minimum angle ($\phi_{0,min}$) for the loop control mirror. Geometrically, the minimum angle can be represented as

\begin{equation}
    \phi_{0,min}=0.5\arctan{\left(\frac{\frac{m}{2}+\frac{\delta_{>99}}{2}}{f}\right)}
    \label{eq:min_angle_1}
\end{equation}

\noindent where $f$ is the focal length of the scan lens, $m$ is the full projected width of the eject mirror, $\delta_{>99}$ is the focal spot diameter within which 99\% of the focused Gaussian beam is contained. This corresponds to a width that is $\approx$ $1.6\times$ the $1/e^2$ Gaussian beam focal spot diameter.

\begin{equation}
    \delta_{>99}=1.6\delta_{1/e^2} = 1.6\biggl(\frac{4}{\pi}\frac{f}{d}\lambda\bigg)
    \label{eq:min_angle_2}
\end{equation}

\noindent where $\lambda$ is the wavelength of light used in ReLOAD and $d$ is the $1/e^2$ collimated beam diameter, and $f$ is the focal length of the scan lens.  In the case where we set the width of the eject mirror to equal to $\delta_{>99}$, 99\% of the beam is ejected on the final pass, and Eq. \ref{eq:min_angle_2} simplifies to

\begin{equation}
    \phi_{0,min}=0.5\arctan{\biggl(\frac{6.4}{\pi}\frac{\lambda}{d}\biggl)}
    \label{eq:min_angle_3}
\end{equation}

\noindent For the demonstration of EO-ReLOAD in this experiment: $d=0.8$ mm, $\lambda=970$ nm, and thus theoretical $phi_{0,min}=1.2$ mrad. Since we decided to configure EO-ReLOAD for only $N=8$ amplification, a substantially larger eject mirror ($m = 1.2$ mm) than the theoretical minimum, and thus larger $phi_{0,min}$ (18 mrad), was still small enough to comfortably fit the loops within $\Phi$.    

\subsection*{Equation 4}
\addcontentsline{toc}{subsection}{Equation 4}

In Eq. 4, we presented an equation for approximating the maximum number of resolvable spots addressable by a deflector in ReLOAD. The number of resolvable spots for a single-pass, based on the FWHM criterion is given as

\begin{equation}
RS_{FWHM}=\left(\frac{FOV}{FWHM}\right)+1
\label{eq:res_spots_supple}
\end{equation}

\noindent where $FOV$ is the field of view, and $FWHM$ is the full width at half maximum of 1D Gaussian focal spot in the scan dimension. Dividing the FOV by $FWHM$ gives the number of spot intervals, and the $+1$ accounts for the two half-spots at the scan endpoints. $FOV$ is given geometrically as 

\begin{equation}
FOV=2f_{KTN}\tan{\theta_{KTN,max}}
\label{eq:FOV_supple}
\end{equation}

\noindent where $\theta_{KTN,max}$ is the maximum deflection half-angle, and $f_{KTN}$ is the effective focal length of the KTN. The $FWHM$ is defined as 0.589 $\times$ the $1/e^2$ spot diameter

\begin{equation}
FWHM=0.589\biggl(\frac{4}{\pi}\frac{f_{KTN}}{d}\lambda\biggl)\approx 0.75\biggl(\frac{f_{KTN}}{d}\lambda\biggl)
\label{eq:fwhm_supple}
\end{equation}

\noindent where $\lambda$ is the wavelength of light in ReLOAD,and $d$ is the $1/e^2$ diameter of the collimated beam. Substituting Eq. \ref{eq:FOV_supple}, and Eq. \ref{eq:fwhm_supple} into Eq. \ref{eq:res_spots_supple} simplifies to the following expression for resolvable spots.

\begin{equation}
RS_{FWHM}=\frac{2}{0.75}\frac{Nd}{\lambda}tan{\theta_{KTN,max}}+1
\end{equation}

\noindent For the demonstration of EO-ReLOAD in this experiment: $N=8$, $d=0.8$ mm, $\lambda=970$ nm, $\theta_{KTN,max}=10$ mrad and thus $RS_{FWHM}=177$ resolvable spots.

\section*{EO-ReLOAD and Performance Characterization Optical Path Details}
\addcontentsline{toc}{section}{EO-ReLOAD and Performance Characterization Optical Path Details}

\subsection*{EO-ReLOAD Details}
\begin{figure}[h!]
\centering
\fbox{\includegraphics[width=.9\linewidth]{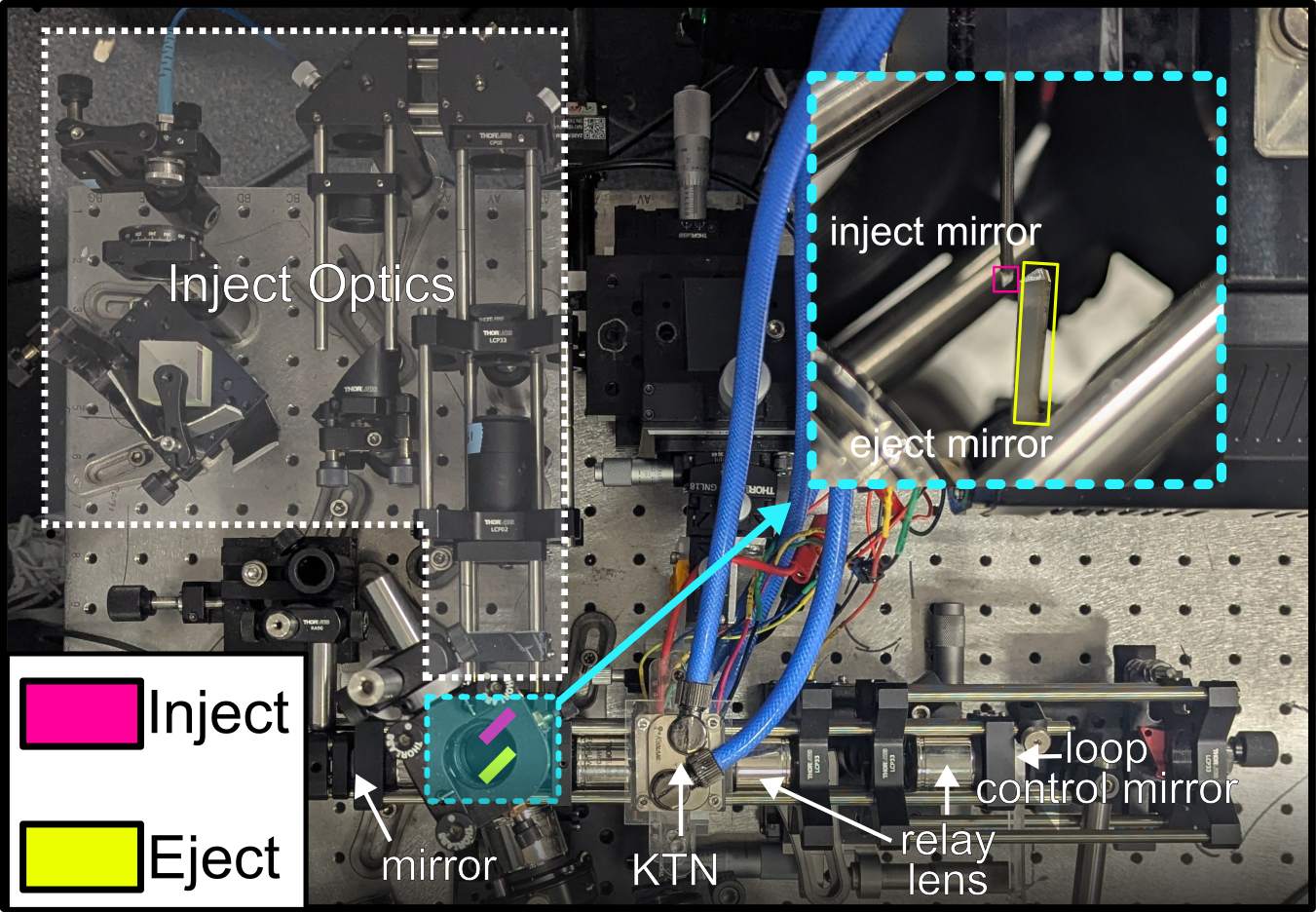}}
\caption{Photograph of EO-Reload implementation.}
\label{fig:trans_reload_image}
\end{figure}

\phantomsection
\addcontentsline{toc}{subsection}{EO-ReLOAD Details}
A photograph of EO-ReLOAD is shown in Figure~\ref{fig:trans_reload_image}. Laser light (Insight X3, Spectra-Physics; 960 nm, 80MHz) was passed through a series of inject optics and focused onto a 1.4 mm $\times$ 1 mm inject mirror (inject M, Edmund Optics, \#86-621), as described in \hyperref[sec:ReLOAD Injection and Output Collection]{ReLOAD Injection and Output Collection} at approximately 4 mm laterally offset from the center of the optical axis along the loop axis (\textit{x}-axis), near the edge of the FOV of a scan lens (LSM03-BB, Thorlabs), which relayed the beam to a conjugate plane at a flat end-mirror (Thorlabs, BB05-E03). After reflection, the beam passed through focus before encountering an identical scan lens that conjugated the end-mirror to the KTN (Fig.~\ref{fig:holder}). The KTN deflected the beam and focused its mode in the scan axis (\textit{y}-axis), and formed half of a cylindrical Galilean telescope together with a plano-concave cylindrical lens custom cut to 10 mm $\times$ 1.2 mm (Thorabs, LK1982L1-B). The deflected beam then exited the KTN/cylindrical assembly both collimated and deflected. A secondary pair of identical scan lenses re-conjugated the KTN pivot plane to a secondary loop control mirror (Thorlabs, BB05-E03) with a slight angle in the loop axis (\textit{x}-axis), reducing the original beam entry angle by 1/4 of the total entry angle. In this way, after four encounters with this end-mirror, the beam trajectory was on-axis when it reflected from the 1.2 mm $\times$ 25 mm eject mirror (Thorlabs, custom cut PFD10-03-P01).

\subsection*{ReLOAD Injection and Output Collection}

\addcontentsline{toc}{subsection}{ReLOAD Injection and Output Collection}
\label{sec:ReLOAD Injection and Output Collection}

\subsubsection*{ReLOAD Inject Details}
\begin{figure}[h!]
\centering
\fbox{\includegraphics[width=.7\linewidth]{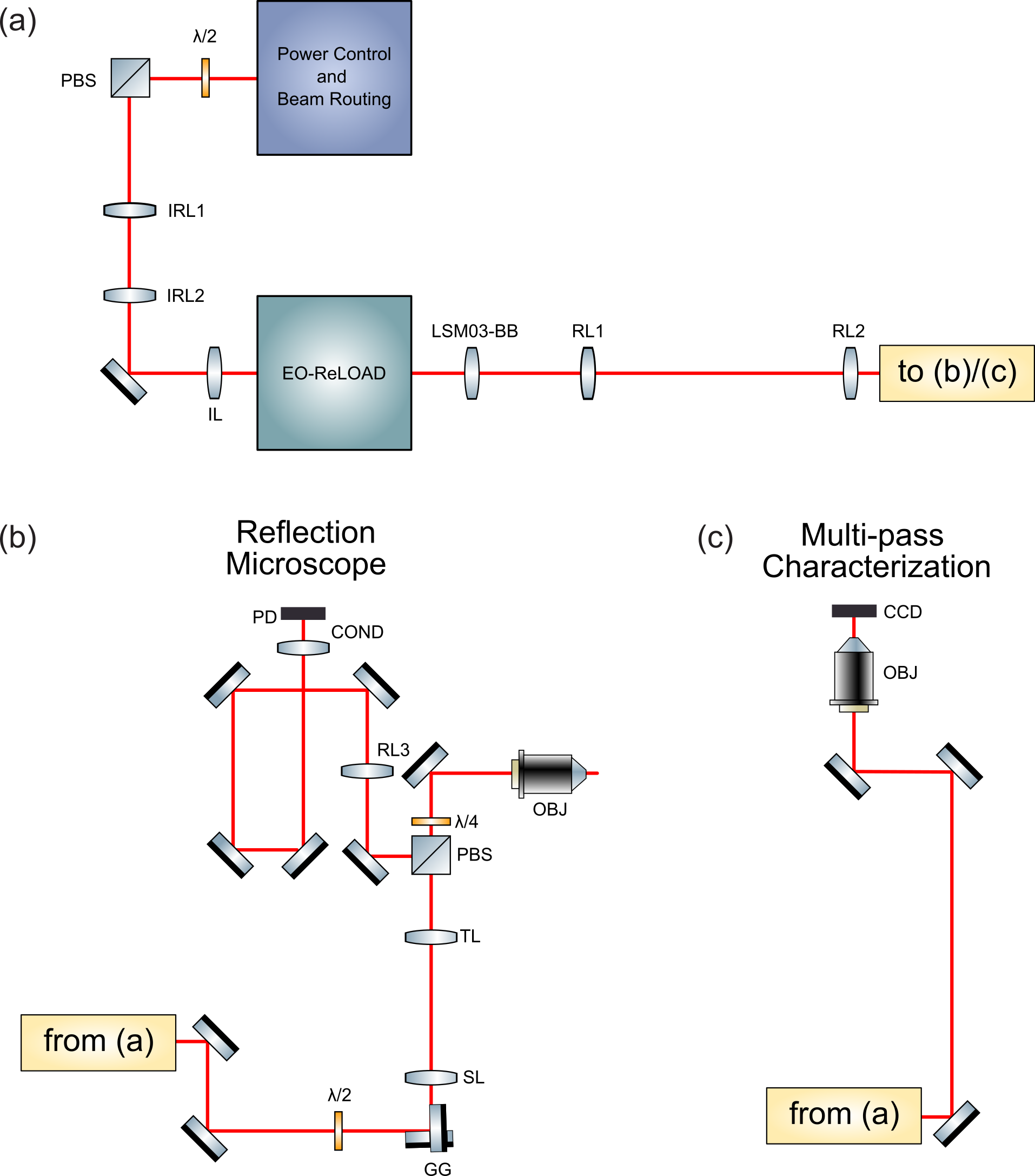}}
\caption{EO-ReLOAD input and output implementation. (a) Layout of the beam path for injection into EO-ReLOAD. (b) Full path through the home built reflectance microscope after exiting EO-ReLOAD, (c) Pick-off path for multi-pass performance characterization.}
\label{fig:mult_ref_scope}
\end{figure}

\phantomsection
\addcontentsline{toc}{subsubsection}{ReLOAD Inject Details}
\label{sec:ReLOAD Inject Details}
The beam path and optical components used for EO-ReLOAD injection are illustrated in Figure~\ref{fig:mult_ref_scope}(a). In this work, EO-ReLOAD was configured for inject beams from either a picosecond pulsed diode laser (Edinburgh Instruments, EPL-980; 970 nm) or a tunable ultrafast source (Insight X3, Spectra-Physics; 960 nm, 80MHz). A $\lambda$/2 waveplate (Thorlabs, WPH05M-980) was used to rotate beams from the EPL-980 or Insight X3 to S-polarization, parallel to the KTN deflection axis. Polarization was further refined by reflection from a polarized beamsplitter cube (PBS), and the beam was transmitted through a telescope composed of IRL1 ($f$ = 125 mm, AC254-125-B) and IRL2 ($f$ = 150 mm, AC254-150-B), producing a collimated 0.8 mm beam. The beam was then focused onto the EO-ReLOAD inject mirror using a $f$ = 40 mm lens (IL, Thorlabs LA1422-B). After ejection from EO-ReLOAD, the beam was routed according to one of the two measurement setups. The Insight X3 beam was sent to a custom reflection microscope, whereas the EPL-980 beam was directed to a CMOS camera for multi-pass beam characterization.

\subsubsection*{Reflection Microscope Setup}
\addcontentsline{toc}{subsubsection}{Reflection Microscope Setup}
\phantomsection
\label{sec:Reflection Microcope Setup}
The reflection microscope setup, including the downstream galvanometer and objective relay, is shown in Figure~\ref{fig:mult_ref_scope}(a,b).  Focal spots were ejected from EO-ReLOAD to a conjugate plane with an external LSM03-BB scan lens. This plane was conjugated to galvanometer mirrors (ScannerMAX Saturn 5b) by a 5$\times$ optical relay consisting of RL1 and RL2. RL1 consisted of three $f$ = 100 mm achromatic doublets and had an effective focal length of $f$ = 33 mm. RL2 consisted of three $f$ = 500 mm achromatic doublets and had an effective focal length of $f$ = 167 mm (Fig.~\ref{fig:mult_ref_scope}(a)). The beam was then relayed and up-magnified by 5.3 $\times$ to fill the 20 mm back-aperture of an objective lens (4$\times$ Nikon Plan Apo $\lambda$) through a custom scan lens (SL) and standard Nikon tube lens (TL). SL consisted of two $f$ = 100 mm achromatic doublets and one $f$ = 150 mm achromatic doublet and had an effective focal length of $f$ = 37.5 mm. Prior to entering the SL-TL relay, the polarization of the laser light was modified to ensure P-polarization prior to passing through a PBS, which was situated near the back-aperture of the objective. After passing through the polarizing beam splitter cube and prior to entering the back aperture of the objective, the laser polarization was modified again by a quarter-wave plate and became circularly polarized before encountering the sample. Reflected light collected by the objective lens passed through this quarter-wave plate again becoming S-polarized. This S-polarized light was then sent to a detection arm by reflection from the PBS. The detection arm relayed the conjugate plane at the back-aperture of the objective to a fast InGaAs photodiode (Thorlabs, DET10C) to resolve the reflected signal from individual laser pulses. This down-magnifying optical relay consisted of a compound relay lens (RL3)  and an aspherical condenser lens (COND) to overfill the detector surface. RL3 consisted of two $f$ = 750 mm achromatic doublets with an effective focal length of $f$ = 375 mm (Fig.~\ref{fig:mult_ref_scope}(b)). All estimated focal lengths of compound lenses are according to the thin lens approximation.

\subsubsection*{Multi-pass Spot Characterization Setup}
\addcontentsline{toc}{subsubsection}{Multi-pass Spot Characterization Setup}
The multi-pass characterization setup is shown in Figure~\ref{fig:mult_ref_scope}(a,c). The EPL-980 beam was ejected from EO-ReLOAD and initially followed the same path as the reflection scope described above (Fig.~\ref{fig:mult_ref_scope}(a)). After passing through the 5× RL1/RL2 optical relay, the beam was picked off and transmitted through the back aperture of an objective lens (4$\times$ Nikon Plan Apo $\lambda$). The beam was then imaged at the focal plane of the objective using a CMOS camera (The Imaging Source, DMK72BUC02), as outlined in \hyperref[sec:Multi and Single-Pass Performance Characterization]{Multi and Single-Pass Performance Characterization} (Fig.~\ref{fig:mult_ref_scope}(c)).

\subsubsection*{Single-Pass Characterization Setup}
\begin{figure}[h!]
\centering
\fbox{\includegraphics[width=.9\linewidth]{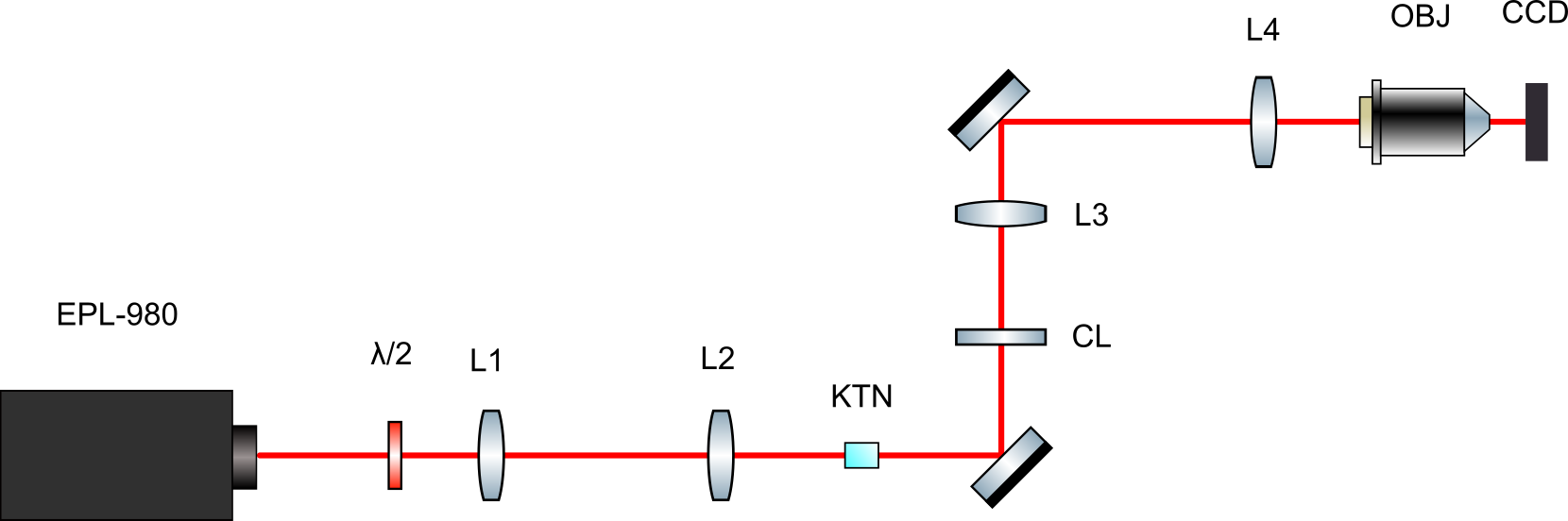}}
\caption{Layout of the beam path for single-pass performance characterization.}
\label{fig:single_pass}
\end{figure}

\addcontentsline{toc}{subsection}{Single-Pass Characterization Setup}
The single-pass characterization set-up is shown in Figure~\ref{fig:single_pass}. An input beam the EPL-980 was rotated with a $\lambda$/2 waveplate (Thorlabs, WPHSM05-980) to be S-polarized, parallel to the KTN deflection axis. The beam was transmitted through a telescope composed of L1 ($f$ = 150 mm, AC254-150-B) and L2 ($f$ = 100 mm, AC254-100-B), producing a collimated 0.8 mm beam. The beam was transmitted through the charged KTN, and the focused beam was re-collimated in the deflection axis with $f$=-30 plano-concave cylindrical lens custom cut to 10 mm $\times$ 12 mm (Thorabs, LK1982L1-B). The beam was then expanded from 0.8 mm to 4 mm using a 5$\times$ telescope composed of L3 ($f$ = 25 mm, AC127-025-B-ML) and L4 ($f$ = 125 mm, AC254-125-B). The enlarged beam was transmitted through the back aperture of an objective lens (4× Nikon Plan Apo $\lambda$) and imaged at the focal plane of the objective using a CMOS camera (The Imaging Source, DMK72BUC02), as outlined in \hyperref[sec:Multi and Single-Pass Performance Characterization]{Multi and Single-Pass Performance Characterization}.

\section*{ReLOAD Optical Design Details}
\addcontentsline{toc}{section}{ReLOAD Zemax Design Details}
\label{sec:ReLOAD Zemax Design Details}

Complete Zemax OpticStudio models of the EO-ReLOAD setup shown in \textbf{Fig. 2}, and reflective mechanical ReLOAD shown in Figure ~\ref{fig:mech_reload} were created. The ray-tracing simulations ensured that the scanning beam could transit each entire system without clipping. This was of particular concern at the EOD scanner aperture. The unfolded sequential designs made use of open source Zemax lens models from Thorlabs, and used periscopes to implement the folding behavior in the loop axis (\textit{x}-axis). In the case of EO-ReLOAD, the KTN EOD was simulated as a Gradient 4 surface with an empirically determined Ny2, where the scanning behavior of the EOD was simulated with an empirically determined \textit{y}-decenter. 

\section*{KTN Crystal Charging and Drive Details}
\addcontentsline{toc}{section}{KTN Crystal Charging and Drive Details}
\phantomsection
\label{sec:KTN Crystal Charging and Drive Details}
The KTN crystal was housed in the custom module described in \hyperref[sec:KTN Module Design]{KTN Module Design}. Both temperature controllers were set to maintain $T_C$ + 5 °C. To charge the crystal and prepare the lensing effect, V+ and V- connections on the module (Fig.~\ref{fig:holder}) were connected to a bridged pair of  DC supplies (GW Instek, GPR-30H10D). 400 VDC was applied for several seconds to ensure full charging. To drive the KTN, a sinusoidal signal from a function generator (FG1, Koolertron, GH-CJDS66-A) was amplified by a pair of high voltage/frequency amplifiers (A1/2, Advanced Energy, Trek 2100HF) bridged with a high-speed phase inverter buffer (PIB, Falco Systems, WMA-IB20). This supplied $\pm$ 296 V at 100 kHz to the KTN. The performance of the amplifiers decreased non-linearly at higher frequencies, resulting in a supply of $\pm$108 V at 500 kHz (Fig.~\ref{fig:fastimaging}). 

\section*{Multi and Single-Pass Performance Characterization}
\phantomsection
\addcontentsline{toc}{section}{Multi and Single-Pass Performance Characterization}
\label{sec:Multi and Single-Pass Performance Characterization}
The EPL-980 laser was triggered by pulses from a function generator (Koolertron, GH-CJDS66-A) synchronized with the KTN drive signal described in \hyperref[sec:KTN Crystal Charging and Drive Details]{KTN Crystal Charging and Drive Details}. Varying the phase of the synchronized pulse signal triggered the laser at specific positions in the EOD scan. After manually determining the phase values corresponding to the endpoints of the scan range, custom Python scripts were used to increment the function generator phase and acquire triggered images of spots across the amplified scan range.
\begin{figure}[h!]
\centering
\fbox{\includegraphics[width=0.75\linewidth]{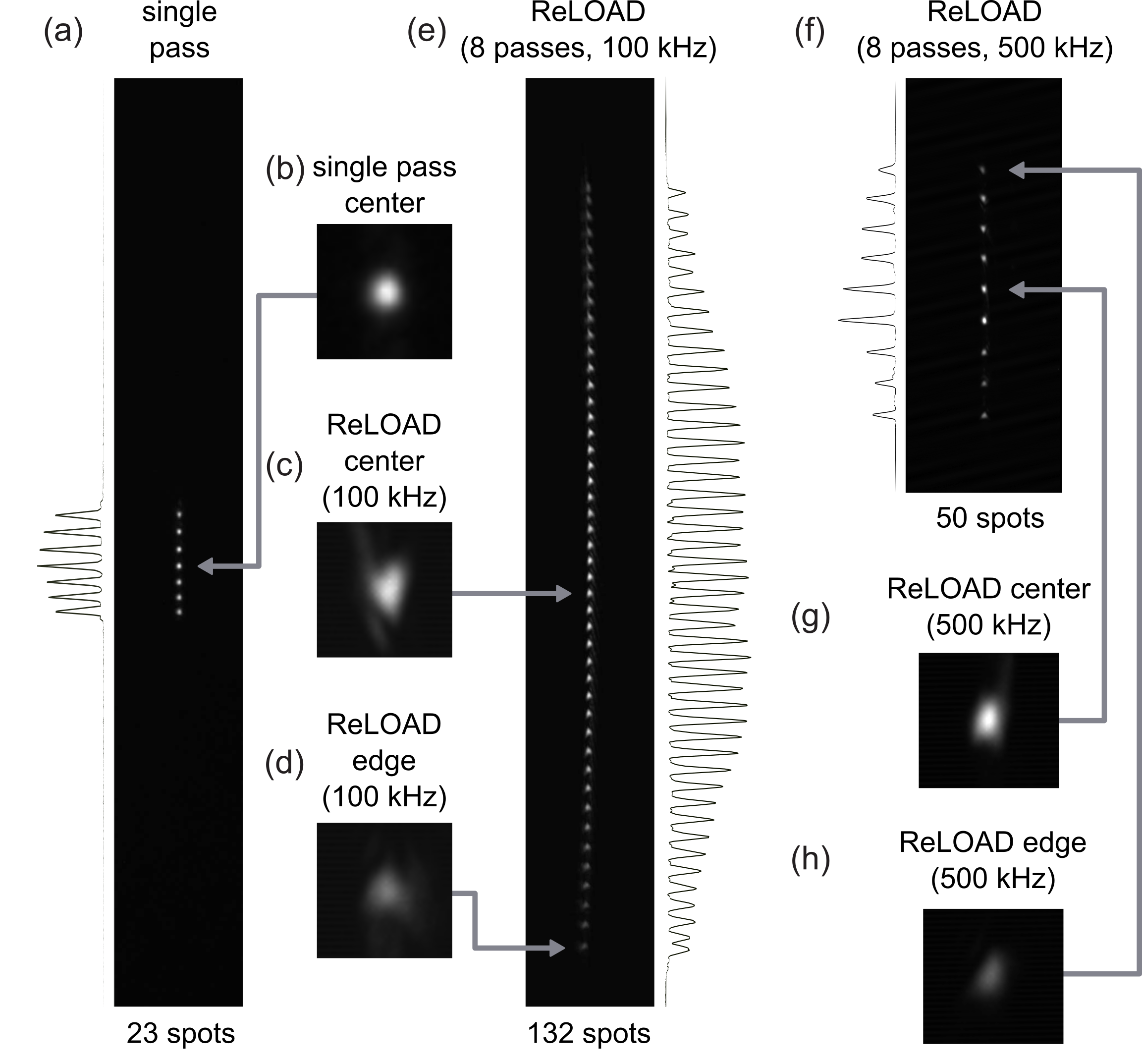}}
\caption{Spots from stroboscopic imaging of EOD and EO-ReLOAD performance. (a) Single pass resolved spots of the KTN deflector. (b) PSF of spot at the center of the single pass line scan. (c) PSF of spot at the center and (d) edge of the EO-ReLOAD at 100 kHz frequency (200 kHz line rate). (e) Amplified resolvable spots recorded after 8 passes through the EO-ReLOAD at 100 kHz frequency. (f) Amplified resolvable spots recorded after 8 passes through the EO-ReLOAD at 500 kHz frequency (1 MHz line rate) . (g) PSF of spot at the center and (h) edge of the EO-ReLOAD at 500 kHz frequency. Lines to the left of (a)/(f) and right of (b) are cross-sections of their associated stroboscopic recordings along the deflection axis.}
\label{fig:all_psf_data}
\end{figure}

In Figure~\ref{fig:all_psf_data}(a), the image was collected stroboscopically by cycling through a set of trigger phases (speed limited by the DDS) corresponding to evenly spaced positions throughout the scan range. The laser power and the CMOS integration time were adjusted to maintain consistent intensity at each spot while avoiding oversaturation at the center of the range. To avoid smearing effects that could occur when using this rapid phase step method over a wider scan range, the images in Figures~\ref{fig:all_psf_data}(e) and ~\ref{fig:all_psf_data}(f) were collected by acquiring individual images at each phase step. The resulting spots were then summed to produce a single image representing an evenly distributed selection of spots across the full scan range. 

In each case, custom Python scripts were used to determine the center of each spot along the deflection axis and fit with a Gaussian profile. The Gaussian fit parameters were then used to calculate the average FWHM of all imaged spots across the range ($\alpha$), as well as the full scan range field of view ($FOV$, center-to-center distance). The number of resolvable spots for a given scan mode was calculated as ($FOV/\alpha$)+1.

\section*{Reflection Microscope Image Acquisition and Processing}
\addcontentsline{toc}{section}{Reflection Microscope Image Acquisition Details}
\subsection*{Image Acquisition}
\begin{figure}[h!]
\centering
\fbox{\includegraphics[width=0.9\linewidth]{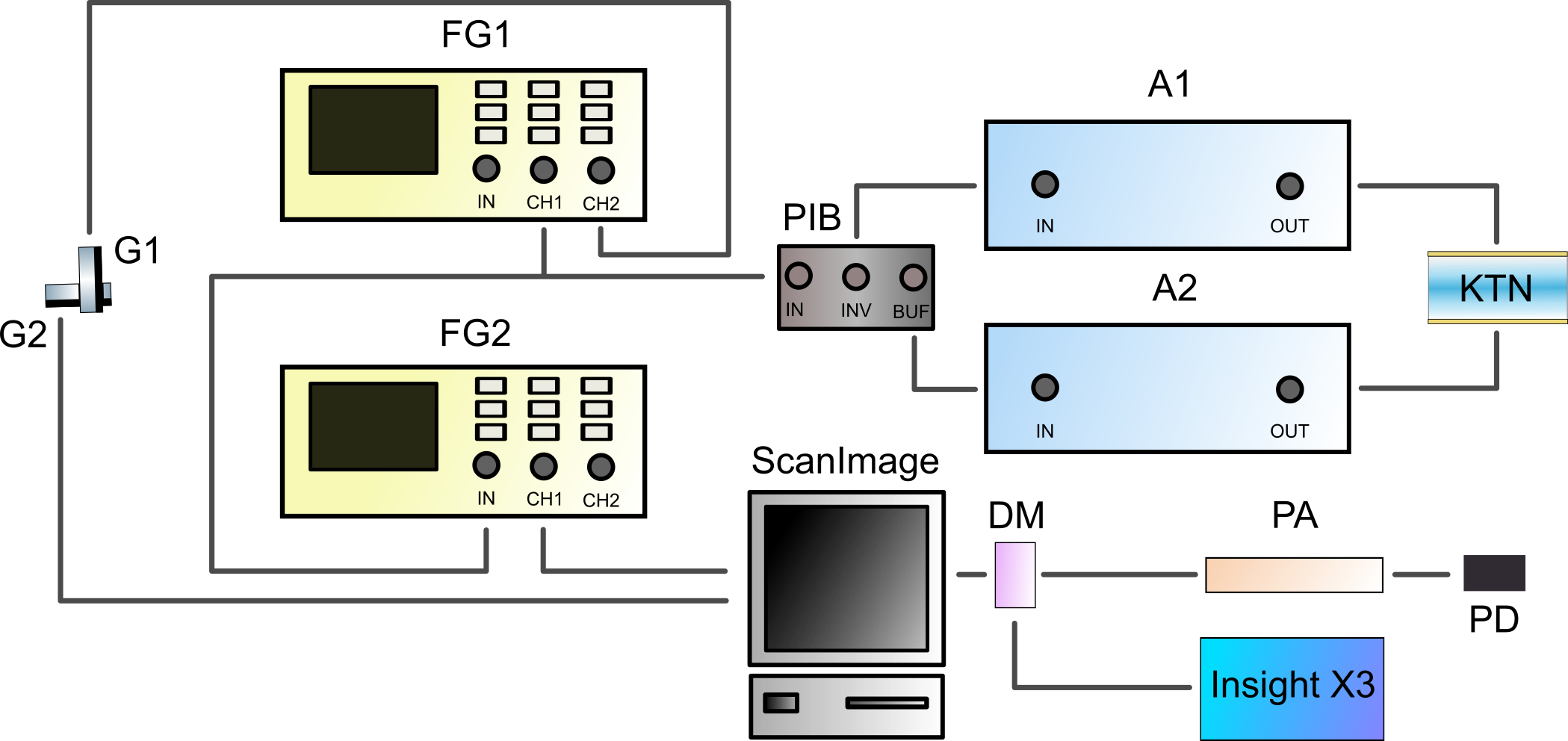}}
\caption{Schematic of electronics used to acquire images in \textbf{Figure 5(a)(b)}. The KTN crystal is charged and driven at 500 kHz (1 MHz linerate) \textit{via} signal generator (FG1, CH1) and bridged amplifiers (PIB, A1/A2), while the signal is simultaneously routed to a second function generator (FG2) to trigger ScanImage acquisition. The photodiode (PD) output is amplified (PA) and digitized by a high-speed module (DM), synchronized to the Insight X3. FG1, CH2 supplies a 5 kHz sinusoid to control the scanning galvanometric mirror (G1), and ScanImage provides control to the non-scanning mirror (G2) for beam-pointing.}
\label{fig:fastimaging}
\end{figure}
\addcontentsline{toc}{subsection}{Image Acquisition}
A reflective chrome University of Minnesota Block M (250 $\mu$m $\times$ 148.75 $\mu$m), etched at the University of Minnesota Nano Center (Heidelberg DWL200), was used as a test sample to characterize the imaging performance of our home-built EO-ReLOAD/galvo reflection microscope. A schematic of the image acquisition setup is shown in Figure \ref{fig:fastimaging}. The KTN crystal was charged with 400 VDC and driven at 500 kHz, as described in \hyperref[sec:KTN Crystal Charging and Drive Details]{KTN Crystal Charging and Drive Details}. The 500 kHz drive signal (FG1, CH1, 1 MHz linerate) was amplified to $\approx \pm$108 V  and sent to the KTN. The signal was simultaneously routed to the input of a second function generator (FG2, IN), which triggered Ch.1 (FG2,CH1) to send pulse signals to an acquisition computer running ScanImage (MBF Bioscience, ScanImage V2023.1.0). The photodiode (PD) output was passed through a preamplifier (PA, Stanford Research Systems, SR445A), which provided a 15× gain, before being digitized by a high-speed module (National Instruments, NI-5771/NI-7975R/PXIe-1092), with sampling synchronized to the repetition rate of the Insight X3 (80 MHz). FG1 Ch.2 supplied a 5 kHz sinusoid to control the scanning galvanometric mirror (G1). The non-scanning mirror (G2) could be controlled in ScanImage to point the beam. Frames were constructed from bidirectional scans along both axes, resulting in imaging at 1 MHz line rate and 10 kHz frame rate. For additional details on the reflection microscope design, see Reflection Microscope Setup and Figure~\ref{fig:mult_ref_scope}(b).
\subsection*{Image Processing}
\addcontentsline{toc}{subsection}{Image Processing}
Movies were processed using custom Python scripts. Non-linear galvanometer and EO-ReLOAD scanning artifacts were sequentially corrected by first approximating the scanning speed using a sine curve generated along the respective axes. These speed curves were then used to construct a pixel-wise remapping using linear interpolation.  Corrections were applied to each frame prior to mean projecting the movies along the temporal axis.

\section*{Funding} National Institutes of Health (RF1NS128658). Fabrication at Minnesota Nano Center: National Science Foundation (ECCS-2025124).

\section*{Acknowledgment} Machining advice and services were provided by the College of Science and Engineering Shop at UMN.

\section*{Disclosures} HJ, CW, DF, SS, AK: University of Minnesota (P). HJ, CW, DF, AK: Arc Photonics (I).

\section*{Data availability}  Data and analysis are available at \href{https://github.com/kerlin-lab/Jayakumar_Warkentin_2025}{www.github.com/kerlin-lab/Jayakumar\_Warkentin\_2025}.

\nolinenumbers
\newpage
\bibliography{ReLOAD_2025_references}
\bibliographystyle{unsrt}
\end{document}